\documentclass[preprint,12pt]{elsarticle}
\usepackage{dcolumn}
\usepackage{graphicx}

\biboptions{sort&compress}

\journal{Annals of Physics}

\begin{document}

\begin{frontmatter}

\title{Comment on: ``Neutron star under homotopy perturbation method'' Ann. Phys. 409 (2019) 167918}

\author{Francisco M. Fern\'{a}ndez}
\ead{fernande@quimica.unlp.edu.ar}

\address{INIFTA, Divisi\'on Qu\'imica Te\'orica,
Blvd. 113 S/N,  Sucursal 4, Casilla de Correo 16, 1900 La Plata,
Argentina}

\begin{abstract}
In this comment we discuss the application of homotopy
perturbation method to a nonlinear differential mass equation that
solves the Tolman-Oppenheimer-Volkoff equation for an isotropic
and spherically symmetric system. We show that one obtains the
same results, more easily and straightforwardly, by means of a
textbook power-series method.
\end{abstract}


\end{frontmatter}

\section{Introduction}

\label{sec:intro}

In a recent paper Aziz et al\cite{ARRG19} derived a mass function that,
according to the authors, provides a solution to the
Tolman-Oppenheimer-Volkoff\cite{T39,OV39} equation for an isotropic and
spherically symmetric system. They solved the nonlinear differential
equation for the mass by means of the homotopy perturbation method and
obtained a seventh-order polynomial function of the radius. With the aid of
Einstein field equations they developed three solutions for different
properties of a neutron star. The purpose of this comment is the discussion
of the application of the homotopy perturbation method to the mass equation.
In section~\ref{sec:mass equation} we derive an approximate solution to that
equation and in section~\ref{sec:conclusions} we summarize the main results
and draw conclusions.

\section{The mass equation}

\label{sec:mass equation}

The core of the paper by Aziz et al\cite{ARRG19} is the nonlinear
differential mass equation
\begin{equation}
m^{\prime }-\frac{1}{2}rm^{\prime \prime }+mm^{\prime \prime }-\frac{5\omega
+1}{2\omega }\frac{mm^{\prime }}{r}-\frac{\omega +1}{2}(m^{\prime })^{2}=0,
\label{eq:mass_eq}
\end{equation}
where $\omega $ is the constant of proportionality between the pressure $p$
and the density $\rho $ in the linear equation of state $p=\omega \rho $.
All the equations are given in units that make the speed of light $c$ and
the gravitational constant $G$ equal to unity. According to the authors,
equation (\ref{sec:mass equation}) solves the Tolman-Oppenheimer-Volkoff one%
\cite{T39,OV39}.

In order to solve equation (\ref{eq:mass_eq}) the authors applied the so
called homotopy perturbation method. It consists of separating the linear
and nonlinear parts of the equation, introducing a perturbation parameter $%
\epsilon $ and expanding the solution in a Taylor series about $\epsilon =0$%
: $m(r)=m_{0}(r)+m_{1}(r)\epsilon +\ldots $. In this way the authors derived
an approximate solution of the form $m(r)=a_{1}r^{3}+a_{2}r^{5}+a_{3}r^{7}$.
The coefficients $a_{j}$ depend on three constants of integration $C_{2}$, $%
C_{3}$ and $C_{5}$ originated in the perturbation equations through second
order. Further analysis shows that $a_{i}\propto C^{i}$, where $C$ is
proportional to the density $\rho _{c}$ at the center of the star ($C=4\pi
\rho _{c}$). The conclusion is that $a_{i}=a_{1}^{i}f_{i}(\omega )$, where $%
f_{i}$ is a rational function of $\omega $. This result looks suspiciously
like a Taylor expansion of $m(r)$ about $r=0$; in what follows we show that
this is actually the case.

For convenience we rewrite equation (\ref{eq:mass_eq}) as
\begin{eqnarray}
D(m) &=&rm^{\prime }-\frac{1}{2}r^{2}m^{\prime \prime }+rmm^{\prime \prime
}-\omega _{1}mm^{\prime }-\omega _{2}r(m^{\prime })^{2}=0,  \nonumber \\
\omega _{1} &=&\frac{5\omega +1}{2\omega },\;\omega _{2}=\frac{\omega +1}{2},
\label{eq:mass_eq_2}
\end{eqnarray}
and look for a solution of the form $m^{[N]}(r)=a_{1}r^{3}+a_{2}r^{5}+\ldots
+a_{N}r^{2N+1}$ such that $D\left( m^{[N]}\right) =\mathcal{O}(r^{2N+3})$.
The calculation of the coefficients $a_{j}$ is trivial and we obtain
\begin{eqnarray}
&&m^{[4]}(r)=a_{1}r^{3}-\frac{3a_{1}^{2}\left( \omega _{1}+3\omega
_{2}-2\right) }{5}r^{5}+\frac{3a_{1}^{3}\left( \omega _{1}+3\omega
_{2}-2\right) \left( 4\omega _{1}+15\omega _{2}-13\right) }{35}r^{7}
\nonumber \\
&&-\frac{a_{1}^{4}\left( \omega _{1}+3\omega _{2}-2\right) \left( 61\omega
_{1}^{2}+2\omega _{1}\left( 243\omega _{2}-224\right) +9\left( 105\omega
_{2}^{2}-192\omega _{2}+88\right) \right) }{315}r^{9},  \label{eq:m^[4]}
\end{eqnarray}
that satisfies $O(m^{[4]})=\mathcal{O}(r^{11})$. We have obtained
one coefficient more than those shown by Aziz et al\cite{ARRG19}
and one can easily derive as many terms as desired without any
effort. Note that they are of the form
$a_{i}=a_{1}^{i}f_{i}(\omega )$ and agree with those obtained by
Aziz et al\cite{ARRG19}, except for the model parameter $n$ that
the authors arbitrarily introduced with the purpose of improving
the accuracy of their theoretical expressions. Without doubt this
approach is far simpler and more straightforward than the homotopy
perturbation method. The form of the coefficients $a_{i}$ shown in
equation (\ref{eq:m^[4]}) suggests that one obtains an exact
solution when $\omega _{1}+3\omega _{2}-2=0$. In fact
\begin{equation}
D\left( m^{[1]}\right) =-3a_{1}^{2}r^{5}\left( \omega _{1}+3\omega
_{2}-2\right) ,
\end{equation}
clearly shows that $m^{[1]}(r)=a_{1}r^{3}$ is an exact solution when $\omega
=-1/3$ or $\omega =-1$, the latter already mentioned by the Aziz et al\cite
{ARRG19}.

\section{Conclusions}

\label{sec:conclusions}

The purpose of this comment is to show that the application of the homotopy
perturbation method to the nonlinear equation (\ref{eq:mass_eq}) in the way
proposed by Aziz et a\cite{ARRG19} is merely a tortuous way of obtaining a
power-series expansion of the solution. The straightforward application of a
textbook power-series method provides much more information with much less
effort. We focus our attention on the solution of equation (\ref{eq:mass_eq}%
) because it appears to be the core of Aziz et al's paper. It is
worth noting that several applications of the homotopy
perturbation method were discussed in the
past\cite{F09,F10,F12,F14} some of which lead to power-series
expansions\cite{F09,F10,F14} and in most of those cases to
extremely poor or even nonsensical results.

\end{document}